\documentclass[english,aps,prb,superscriptaddress,nobibnotes,amsmath,amssymb,showkeys,floatfix]{revtex4-1}

\usepackage[utf8]{inputenc}
\usepackage[english]{babel}
\usepackage{amsmath,amsfonts,amssymb}
\usepackage{upgreek}
\usepackage[T1]{fontenc}
\usepackage{url}
\usepackage{placeins}

\usepackage{graphicx}

\usepackage{titlesec}
\titleformat{\section}{\normalfont\large\bfseries\filcenter}{\thesection}{1em}{}
\titlespacing{\section}{0pt}{10pt}{5pt}

\begin{document}

\title{From the Lugiato-Lefever equation to microresonator based soliton Kerr frequency combs}

\author{L.A. Lugiato}
\affiliation{Dipartimento di Scienza e Alta Tecnologia, Universit\`{a} dell'Insubria, via Valleggio 11, 22100 Como, Italy}
\author{F. Prati}
\affiliation{Dipartimento di Scienza e Alta Tecnologia, Universit\`{a} dell'Insubria, via Valleggio 11, 22100 Como, Italy}
\email{franco.prati@uninsubria.it}
\author{M.L. Gorodetsky}
\affiliation{Faculty of Physics, Lomonosov Moscow State University, 119991, Moscow, Russia}
\affiliation{Russian Quantum Center, 143025 Skolkovo, Russia}
\author{T.J. Kippenberg}
\affiliation{\'Ecole Polytechnique F\'ed\'erale de Lausanne (EPFL), Lausanne, CH-1015, Switzerland}


\begin{abstract}
The model, that is usually called Lugiato-Lefever equation (LLE), was introduced in 1987 with the aim of providing a paradigm for dissipative structure and pattern formation in nonlinear optics. This model, describing a driven, detuned and damped nonlinear Schroedinger equation, gives rise to dissipative spatial and temporal solitons. Recently, the rather idealized conditions, assumed in the LLE, have materialized in the form of continuous wave driven optical microresonators, with the discovery of temporal dissipative Kerr solitons (DKS). These experiments have revealed that the LLE is a perfect and exact description of Kerr frequency combs - first observed in 2007, i.e. 20 years after the original formulation of the LLE. -  and in particular describe soliton states.
Observed to spontaneously form in Kerr frequency combs in crystalline microresonators in 2013,
such DKS are preferred state of operation, offering coherent and broadband optical frequency combs, whose bandwidth can be extended exploiting soliton induced broadening phenomena. Combined with the ability to miniaturize and integrate on chip, microresonator based soliton Kerr frequency combs have already found applications in self-referenced frequency combs, dual-comb spectroscopy, frequency synthesis, low noise microwave generation, laser frequency ranging, and astrophysical spectrometer calibration, and have the potential to make comb technology ubiquitous. As such,  pattern formation in driven, dissipative nonlinear optical systems is becoming the central Physics of soliton micro-comb technology.

\end{abstract}

\maketitle

\section{Introduction}\label{sec:1}

One of the most fascinating classes of phenomena that arise in dissipative systems is the spontaneous gradual formation of stationary spatial patterns from initial states that are spatially uniform. With the introduction of the general concept of dissipative structure, Ilya Prigogine showed that such phenomena are ubiquitous in far from equilibrium nonlinear open systems, whenever the onset of a Turing instability \cite{Turing52} triggers the self-organization of the system \cite{Prigogine77}. Thus, one meets this scenario in  nonlinear dynamical systems of  a wide  variety of fields, encompassing e.g. chemistry, hydrodynamics, biology , population dynamics, social sciences (see in this connection also Haken's Synergetics \cite{Haken77}). 

In 1977 one of us (LL) spent a few months as a guest of Prigogine's group. The book of Nicolis and Prigogine \cite{Prigogine77} had just appeared, and this circumstance triggered a natural wish of formulating, in the framework of nonlinear optics/photonics, a model that could play the same role of the Prigogine-Lefever model \cite{Prigogine68}, usually designated with the name Brusselator, in the framework of nonlinear chemical reactions. 

The case of optics is especially interesting in the vast domain of nonlinear dynamical systems. Prigogine himself underlined that in optical systems the nonlinear dynamics is governed by the fundamental laws of radiation-matter interaction. Other two main reasons  are the following \cite{Lugiato94}. First, optical systems can exhibit a fast dynamics and can operate over very large frequency bandwidths, and these features have led to relevant practical applications, especially in telecommunications.
The second reason is that they can exhibit relevant  quantum effects, without need for reducing the temperature to extremely low values. Thus, they lend themselves naturally for the implementation of quantum technologies. 

The aim of formulating an optical model for pattern formation, with the same level of paradigmatic simplicity of the Brusselator,  materialized in 1987, with the introduction of the equation formulated in collaboration with Ren\`{e} Lefever \cite{LLE}. In the following we will indicate it as the LLE. 

From a mathematical viewpoint, the LLE amounts to a damped, driven nonlinear Schr\"odinger equation with detuning.
From a physical standpoint, the main criterion followed to construct the LLE was that of maximum simplicity. Thus the choice of the nonlinearity led to the selection of the cubic Kerr nonlinearity, which allows for formulating a self-contained equation for the slowly varying envelope of the electric field only. The Kerr medium is assumed to be contained in a high-Q  ring cavity driven by a continuous wave (CW) field. The use of the low transmission limit \cite{NLOS} allows for formulating a model which does not involve the spatial longitudinal variable $z$ along which light propagates. Hence the original model introduced in \cite{LLE} describes the spontaneous onset of stationary patterns in the transverse planes $x-y$, orthogonal to the propagation direction. In this case pattern formation arises from the combination of the Kerr nonlinearity and of diffraction of radiation, similarly to the case of  the Brusselator, in which pattern formation is produced by the combination of the nonlinearity and of diffusion of the two chemicals in play. 2D patterns involve both transverse variables $x$ and $y$, whereas 1D patterns involve only one transverse variable, say $y$. An especially important class of patterns is given by cavity solitons \cite{NLOS}.

Some years after the introduction of the LLE, Haelterman et al. \cite{Haelterman92} formulated the temporal/longitudinal version of the LLE. The main step to do that was to replace diffraction by group velocity dispersion, i.e. to substitute the Laplacian with respect to the transverse variables $x$ and $y$, which describes diffraction, by the second derivative with respect to the retarded time in the cavity, which describes dispersion. Thus the LLE of \cite{Haelterman92} involves as independent variables the time $t$ and the retarded time $\tau=t-z/v_g$, where $v_g$ is the group velocity. Hence it describes the spontaneous formation of spatial patterns (and cavity solitons) in the longitudinal direction $z$, while the field envelope remains uniform in the transverse plane $x-y$, whereas in the original LLE of \cite{LLE} patterns and solitons develop in the transverse plane, and the field envelope remains uniform in the $z$ direction of the cavity. The longitudinal patterns of \cite{Haelterman92} propagate along the cavity with velocity $v_g$, and for this reason they are temporal/longitudinal patterns. 

From a mathematical viewpoint, however, the temporal/longitudinal version of the LLE is fully equivalent to the transverse LLE in 1D.

Two relevant remarks are in order. The first is that the work \cite{SUSSP93} formulated a longitudinal LLE equivalent to that of \cite{Haelterman92}, deriving it from the Maxwell-Bloch equations (MBE) \cite{NLOS} in the dispersive limit. The detailed derivation can be found in \cite{Castelli17}.

The second remark is related to the topic of optical bistability. It is well known that a driven cavity containing a Kerr medium exhibits bistability \cite{Gibbs76}. Phenomena of spontaneous pattern formation along a ring cavity driven by a cw field, with the same scenario offered by the longitudinal LLE, have been previously studied  in the framework of the multimode instability of optical bistability predicted in \cite{Bonifacio78,Bonifacio79} by using  the MBE, and experimentally observed later in Ref. \cite{Segard89}. Hence, on the basis of the analysis of Refs. \cite{SUSSP93} and \cite{Castelli17} we see that the instability of the longitudinal LLE model can be considered as a special case of the multimode instability of optical bistability. However, in the parametric conditions which characterize the longitudinal LLE, the number of longitudinal cavity modes activated by the instability is typically much larger than in the case of Refs. \cite{Bonifacio78,Bonifacio79}, a feature which is important with respect to the phenomenon of broadband Kerr frequency combs that we are going to address.


Let us now make a temporal jump to the new century. Optical frequency combs constitute an equidistant set of laser frequencies that can be employed to count the cycles of light, a technique introduced and developed by Theodor H\"ansch and John Hall. Their realization using mode-locked lasers \cite{Jones00,Udem02} introduced a revolution in the ability to measure optical frequencies, and enabled optical atomic clocks, precision measurements of fundamental constants, as well as advances in many other areas including spectroscopy, distance measurements or time transfer.
In 2007 Del'Haye et al. \cite{DelHaye07} demonstrated the realization of broadband optical frequency combs exploiting the whispering gallery modes activated by a CW field driving a high-Q microresonator containing a Kerr medium. This result broke with the conventional thinking that combs can only be generated by a mode locked pulsed laser source (as exemplified e.g. by the anonymous peer reviewer expressing: "In these Kerr-combs, there is no mechanism to force the comb lines to be perfectly spaced, unlike a mode-locked laser where the mode-locking mechanism forces the lines to be phase locked to one another"). In the present case, the broadband optical frequency comb is generated by the complex four-wave mixing  (FWM) processes induced by the interaction between the monochromatic driving field and the Kerr medium. 
The possibility to synthesize microresonator based frequency combs whose bandwidth can exceed an octave with repetition rates in the microwave to THz domains offers substantial potential for miniaturization and chip-scale photonic integration, as well as power reduction  offers access.  These characteristics have led to substantial research activity. 
Today, Kerr frequency comb generation is a mature field, and the technology has been applied to numerous areas, including coherent telecommunications, spectroscopy, atomic clocks as well as laser ranging and astrophysical spectrometer calibration. A key impetus to these development have been dissipative Kerr solitons, which are low noise operating states in Kerr combs, offering coherence, and the ability to achieve substantial bandwidth via soliton induced Cherenkov radiation \cite{Brasch16,Akhmediev95}.

The link between the topic of Kerr optical frequency comb generation and the mean field LLE was theoretically first developed by Matsko \cite{Maleki10,Matsko11}. While the majority of early work in the field focused on the frequency domain representation of four wave mixing via coupled mode equation, this work introduced a mean field model in the temporal domain. 
This important theoretical insight did not lead directly to new experimental advances as neither exact conditions for soliton formation in microresonators were  known, nor experimental evidence were observed. First evidence for a new operation regime appeared in 2012, when it was observed that in microresonators made from $\mathrm{Si_3N_4}$ a transition from state containing multiple subcombs to a low phase noise state is possible \cite{Herr12}. Using crystalline resonators, single (and multiple) dissipative Kerr solitons were unambiguously and stably accessed in a crystalline microresonators for the first time in  2014.
The link with correct expression for the soliton parameters was first given in 2012 in the arXiv version \cite{Soliton_Arxiv} of paper \cite{Herr14}. The derivation was based on a paper by Wabnitz \cite{Wabnitz93}, who utilized  earlier results obtained by Nozaki with Bekki \cite{Nozaki84} and by Kaup with Newell \cite{Kaup78} for the damped, driven nonlinear Schr\"odinger equation in the framework of plasma physics. Though in \cite{Matsko11} the link appeared possibly the earliest, the crucial importance of the \emph{effective laser pump detuning} as a free parameter was not recognized there.
Later work noted the rigorous connection to the LLE, in particular the work by Chembo et al. \cite{Chembo13,Godey14}, and Coen et al. \cite{Coen13}.

These authors showed that the LLE (or generalizations including higher order dispersion terms) is the model which describes the generation of Kerr frequency combs and is capable of predicting the properties of Kerr combs when the system parameters, notably the laser detuning, are varied. The spontaneous formation of spatial patterns traveling along the cavity, described by the LLE, is the spatiotemporal equivalent of the frequency combs and governs their features. The rather idealized conditions assumed in the formulation of the LLE have been perfectly materialized by the spectacular technological progress that has occurred in the meantime  in the field of photonics and has led, in particular, to the discovery of Kerr frequency combs.  
A surprising and beautiful conspiracy is that while thermal effects make the access to the soliton state unstable, the same thermal effect stabilizes the soliton state indefinitely once they are accessed.    

Experimentally observed combs have been compared with simulations by the LLE e.g. in \cite{Coen13,Herr12,Lamont13,Coen13a,Grudinin17}. The analytic expression for the soliton formulated in \cite{Soliton_Arxiv} has been proven capable of approximating very well both numerical simulations of the LLE and experimental data; the special case of this expression, in which the background of the soliton is neglected, was utilized later in \cite{Coen13a}.


The mechanism that lies at the root of the multimode instability of optical bistability, as well as of the instabilities of the longitudinal LLE model, is that under appropriate parametric conditions the cw driving field can create gain in longitudinal cavity modes different from the one with which the injected field is quasi-resonant. Hence, even if the system is passive, in the sense that it does not display population inversion, the driving field induces gain in side-modes of the resonant mode, and the gain originates the instability, the traveling pattern and the pulsed output. Therefore the system is definitely active with respect to the side-modes, and represents a source. 

Reviews of the field of Kerr combs can be found in \cite{Kippenberg11,Herr15,Savchenkov16,Chembo16,Pasquazi17}.

In Sec.2 we describe the long history of the LLE, adding all the relevant details to what is already discussed in the introduction. 
In Sec.3 the experimental development of the Kerr frequency combs towards solitons is described. 
Sec. 4 discusses dissipative Kerr solitons in photonic integrated microresonators.
Some conclusions are drawn in the final section.

\section{Motivation and history}\label{sec:2}

\subsection{The formulation of the LLE}\label{sec:2a}

The Brusselator model \cite{Prigogine68} consists in two coupled nonlinear equations that describe the interaction of two reactants in an open environment. The joint action of nonlinearity and diffusion produces the formation of Turing patterns, which are described in 2D, because the system is contained in a vessel which has large dimensions in the directions $x$ and $y$, and negligible dimension in the direction $z$, so that the variable $z$ does not appear in the model.

\begin{figure}[ht]
\centering\includegraphics[width=0.8\linewidth]{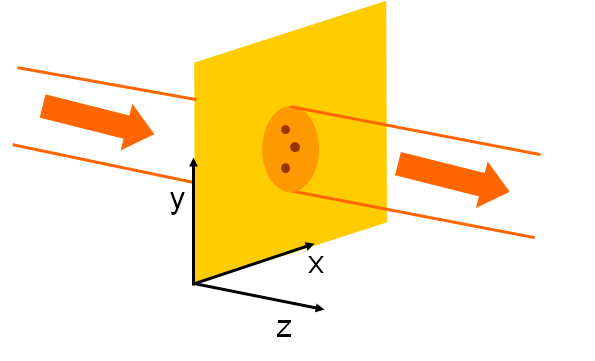}
\caption{A transverse pattern may arise when a broad section coherent beam interacts with a nonlinear medium.}
\label{fig:1}
\end{figure}

In the case of optics, the coordinates $x$ and $y$ are those which span the planes orthogonal to the propagation direction $z$ (see Fig. \ref{fig:1}). 
The role of diffusion is played by diffraction which, in the par-axial approximation, is described by a term proportional to the transverse Laplacian (multiplied by the imaginary unit) of the electric field envelope, while diffusion is described by terms proportional to the transverse Laplacian of the concentration of the reactants. The complex field envelope $E(x,y,z,t)$, where $t$ denotes time, is related to the electric field (assumed linearly polarized) ${\cal E}(x,y,z,t)$ as it follows:
\begin{equation}\label{eq:1}
{\cal E}(x,y,z,t) = \frac{1}{2}E(x,y,z,t) \exp[-i \omega_p (t-z/c)]  +\mathrm{c.c.} \,,
\end{equation}
where $\omega_p$ is the frequency of the input field and $c$ is the light velocity in the medium.

The starting point for the formulation of the LLE was the system of partial differential equations corresponding to the MBE \cite{NLOS}, which describe the dynamics of the electric field envelope coupled to that of a system of two-level atoms. In order to reduce to a set of two equations like the Brusselator, one must assume parametric conditions that allow for the adiabatic elimination of the atomic variables. In this way one obtains a self-contained equation for the complex field envelope $E$, which amounts to two real equations. 
However, a glance at Fig. \ref{fig:1} shows that in optical systems the laser field propagates in the longitudinal direction $z$ which, therefore, cannot be ignored. Let us consider, for the sake of definiteness, the case that the nonlinear medium fills a ring cavity. The field propagates in the cavity with a characteristic time, the cavity roundtrip time ${\cal L}/c$, where ${\cal L}$ is the length of the ring, and this time is the inverse of the free spectral range of the cavity. The propagation of the field must be calculated for a large number of roundtrips and, even if the dynamics can be reduced to an Ikeda map \cite{Ikeda79}, the first models which were used to describe transverse diffractive pattern formation \cite{Moloney82,McLaughlin83} were far from the simplicity of the Brusselator. 

The way out to solve this problem is provided by the so-called low transmission limit or high-quality (high-Q) cavity limit, which is described in detail, for example, in Refs. \cite{NLOS} and \cite{Castelli17}. If, in addition to this limit, one assumes that only the cavity mode that is closest to resonance with the input field has a nonzero amplitude (singlemode limit), one has that the field envelope becomes uniform along the cavity, so that it varies only with respect to time and to the transverse variables $x$ and $y$, while the longitudinal variable $z$ becomes irrelevant as in the Brusselator. Contextually, in this limit the time scale of the dynamics is no longer the cavity roundtrip time, but the cavity decay time ${\cal L}/cT$, where $T$ is the transmissivity coefficient of the cavity mirrors which is very small in the low transmission limit, so that the cavity decay time is much longer that the roundtrip time. This longer time coincides with the inverse of the cavity linewidth.

The adiabatic elimination of the atomic variables leads to a saturable nonlinearity, in which the nonlinear term has the form $E f(|E|^2)$, where the function $f$ can be expanded in Taylor series of $|E|^2$ and includes all even powers of $|E|^2$. Here, continuing to follow the criterion of maximum simplicity, one keeps only the linear term in $|E|^2$, so that the nonlinear term that appears in the equation has the cubic form $E|E|^2$, that is typical of Kerr media. In conclusion the LLE, derived in the way we have described above, involves the following terms: the time derivative, the transverse Laplacian which describes diffraction, the Kerr nonlinear term, a term which describes the driving input field and two terms related to the ring cavity:
\begin{equation}\label{eq:2}
\frac{\partial E}{\partial t}=F-E-i\alpha E+i|E|^2E+i\nabla^2_\perp E\,, \qquad 
\nabla^2_\perp=\frac{\partial^2}{\partial x^2}+\frac{\partial^2}{\partial y^2}\,.
\end{equation}
\begin{figure}[ht]
\centering\includegraphics[width=0.8\linewidth]{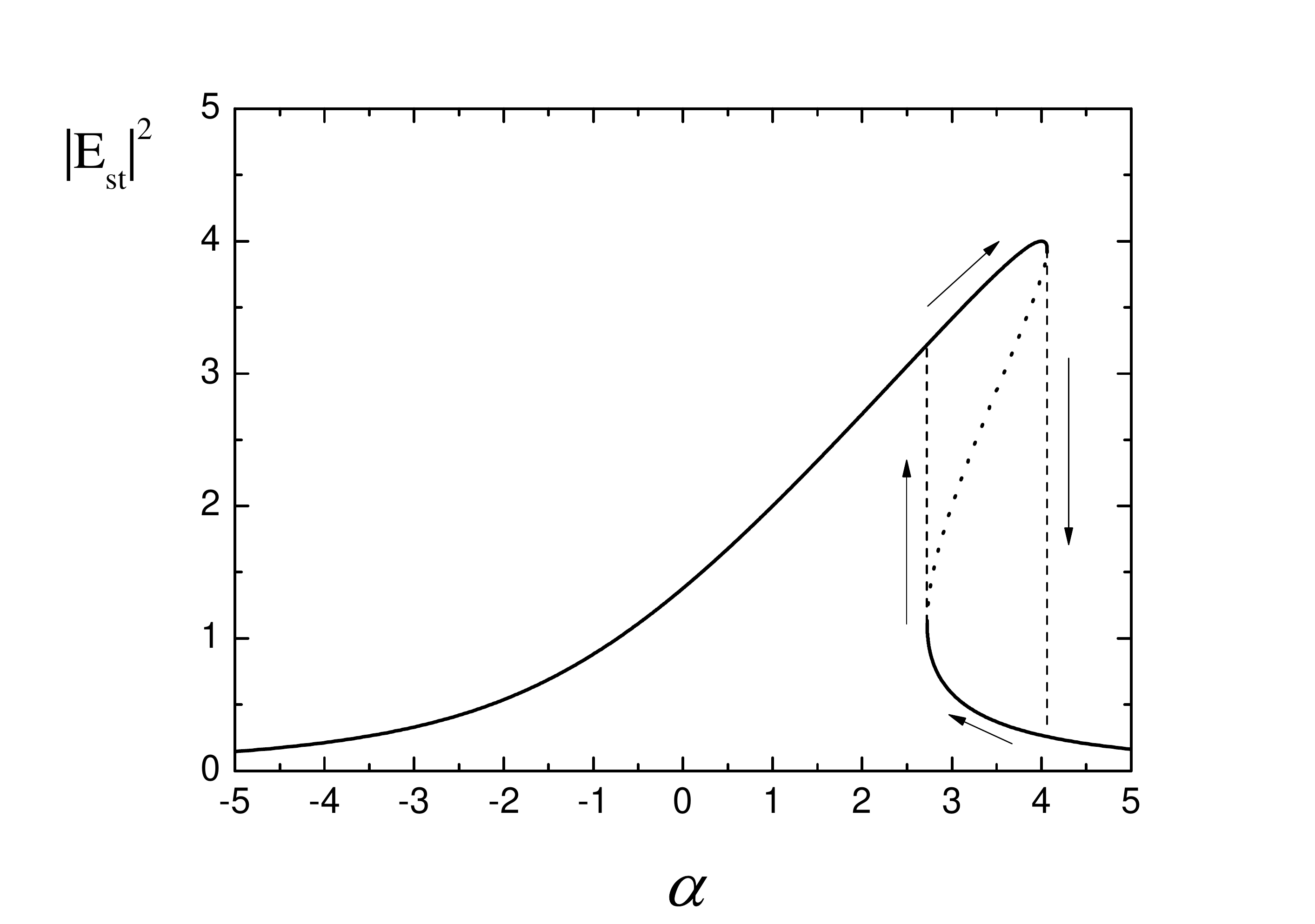}
\caption{Plot of the stationary curve (\ref{eq:5}) for $F^2=4$. The broken line with negative slope is unstable. When the detuning parameter $\alpha$ is varied, the system covers a hysteresis cycle.}
\label{fig:2}
\end{figure}

In Eq. (\ref{eq:2}) $E$ and $F$ (the input field amplitude) are appropriately normalized in order to reduce to the minimum the number of parameters which appear in the model \cite{LLE,NLOS}. The time $t$ is normalized to the cavity decay time, the spatial variables $x$ and $y$ are normalized to the diffraction length in the cavity \cite{Castelli17}. We have assumed to be in the self-focusing case. In the r.h.s of Eq. (\ref{eq:2}) the first term introduces the input field, the second describes the escape of photons from the cavity, the third is a detuning term, with 
$\alpha=\frac{\omega_0-\omega_p}{cT/{\cal L}}$ where $\omega_0$ is the cavity frequency closest to the input frequency $\omega_p$. Assuming that the input field $F$ is real, the homogeneous ($\nabla^2_\perp E=0$) stationary ($\partial E/\partial t=0$) solutions of Eq. (\ref{eq:2}) obey the cubic equation
\begin{equation}\label{eq:5}
F^2=|E|^2\left[1+\left(\alpha-|E|^2\right)^2\right]\,.
\end{equation}
Figure \ref{fig:2} shows the stationary curve of the normalized transmitted intensity $|E|^2$ as a function of $\alpha$ for the fixed value $F^2=4$ of the normalized input intensity. There is a range of values of $\alpha$ for which one has three stationary solutions, but the intermediate solution which lies in the negative slope portion of the curve is unstable. Therefore in that range the system is bistable and, if the value of the detuning parameter is swept forth and back for fixed input intensity, one covers a hysteresis cycle.

The linear stability analysis of the homogeneous stationary solutions \cite{LLE,SNS87} demonstrates that under appropriate parametric conditions, one or more segments of the stationary curve with positive slope are unstable, and this circumstance opens the possibility of the spontaneous formation of stable stationary Turing patterns. Let us focus first on the 1D case of one transverse direction, neglecting the transverse variable $x$ and replacing the transverse Laplacian in Eq. (\ref{eq:2}) by the second derivative with respect to $y$.

The Kerr nonlinearity corresponds to a process of FWM. A simple possibility is that two input photons injected in the longitudinal direction $z$ get absorbed by the medium, and simultaneously the system emits two photons which propagate in symmetrically tilted directions 
with transverse component of the wave vector equal to $k$ and $-k$, respectively. Thus the envelope $E$ takes the typical sinusoidal configuration
\begin{equation}\label{eq:6}
E(y)=E_{st}+\sigma\mathrm{e}^{i\phi_+}\mathrm{e}^{iky}+\sigma\mathrm{e}^{i\phi_-}\mathrm{e}^{-iky}
=E_{st}+2\sigma\cos\left(ky+\frac{\phi_+-\phi_-}{2}\right)\exp\left(i\frac{\phi_++\phi_-}{2}\right)\,,
\end{equation}
where $E_{st}$ denotes a homogeneous stationary solution.
A circumstance of outstanding importance is that the two symmetrically tilted photons are in a state of quantum entanglement (for example their energy and momenta are strongly correlated at the quantum level). The difference between the number of photons emitted in the two directions is squeezed, i.e. exhibits fluctuations below the shot noise level. This effect, theoretically predicted in \cite{Castelli92}, has been recently observed experimentally in a Kerr ring microcavity \cite{Dutt15}, in a longitudinal configuration corresponding to the temporal/longitudinal LLE that will be discussed in subsection \ref{sec:2c} of this section. This kind of quantum aspects are fundamental for the field of quantum imaging \cite{Gatti08,Kolobov99}.

\subsection{2D patterns and spatial cavity solitons}\label{sec:2b}

\begin{figure}[ht]
\centering\includegraphics[width=1.0\linewidth]{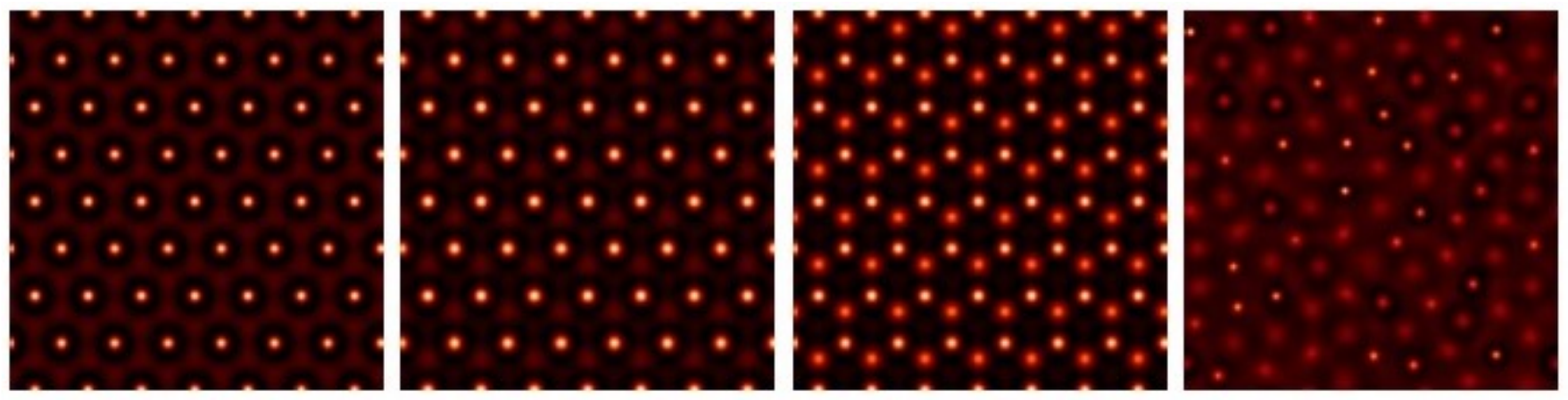}
\caption{Time evolution of the field of an unstable hexagonal pattern. The time increases from left to right. At the end one has a chaotic situation. Reprinted figure with permission from \cite{Scroggie94}. Copyright (2007) by the American Physical Society.}
\label{fig:4}
\end{figure}

Let us now turn to the 2D configuration considering both transverse variables $x$ and $y$. In this case Eq. (\ref{eq:6}) corresponds to a roll (i.e. stripe) pattern. However, as shown in \cite{Grynberg88}, in 2D the roll pattern is unstable, and further FWM processes give rise to a hexagonal pattern. References \cite{Gomila03,Gomila07} analyzed the scenario of hexagonal structures that arises as parameters are varied; in many cases the pattern exhibits a dynamical (chaotic) behavior (see Fig. \ref{fig:4})

\begin{figure}[ht]
\centering\includegraphics[width=0.8\linewidth]{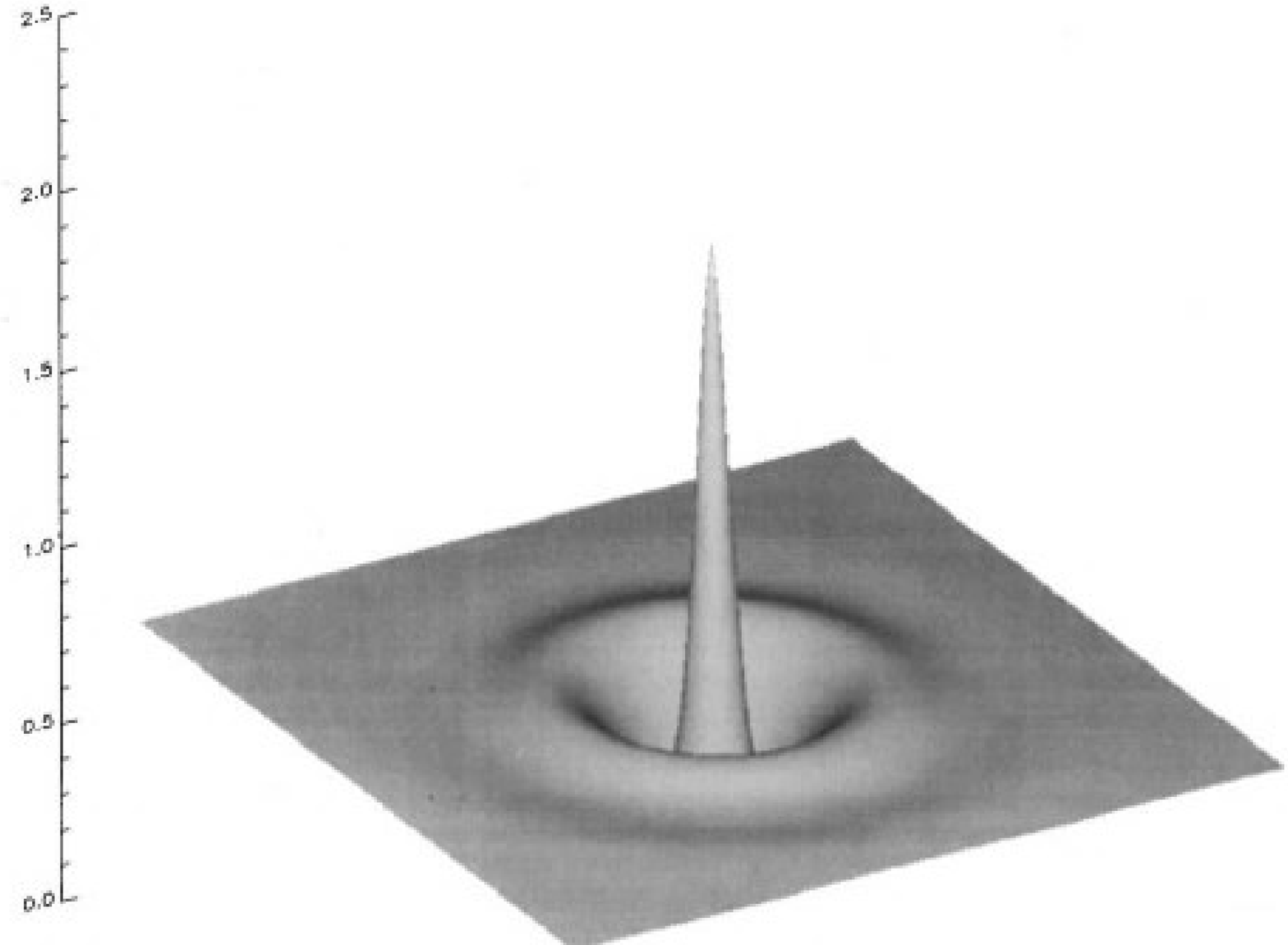}
\caption{A typical Kerr cavity soliton, showing a bright peak
on a darker homogeneous background with a few weak diffraction
rings. Reprinted figure from reference \cite{Firth02} with permission from the Optical Society of America.}
\label{fig:5}
\end{figure}

In the general field of spatial pattern formation, one usually considers patterns which are formed by elements well correlated to one another. However, one meets also the case of localized structures, formed by one or more elements that are independent provided that they are not too close to one another \cite{Coullet04}. In the framework of nonlinear optics localized structures were first predicted by Tlidi et al. in \cite{Tlidi94}, and they are usually called spatial cavity solitons. This kind of solitons were analyzed in the framework of the LLE in \cite{Firth02,Scroggie94} (see Fig. \ref{fig:5}). In \cite{Scroggie94} it is also shown that, when the input intensity is increased, the soliton starts breathing, i.e. its peak intensity oscillates in time. Reviews on spatial cavity solitons can be found in \cite{Lugiato03,Ackemann09}.

\begin{figure}[ht]
\centering\includegraphics[width=0.8\linewidth]{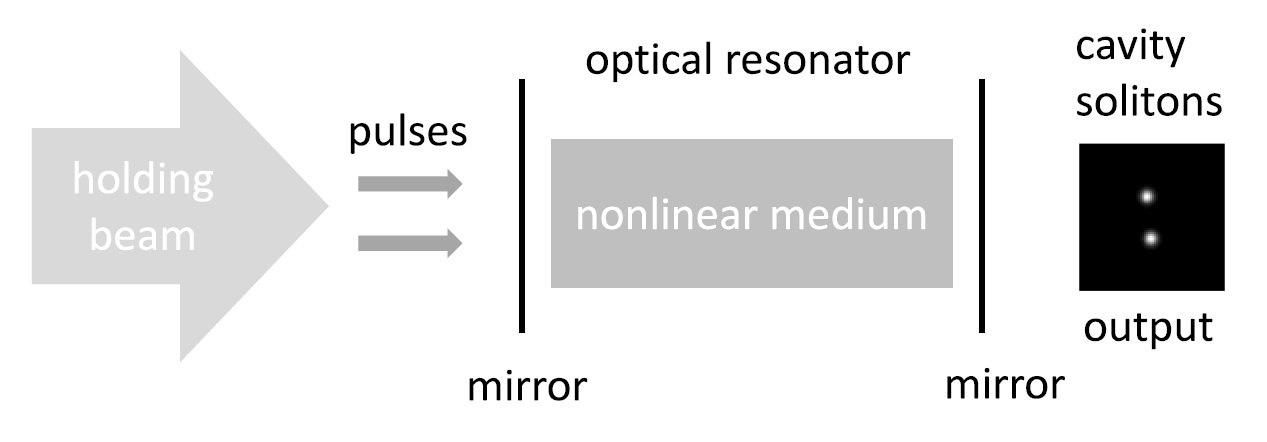}
\caption{A coherent, stationary, quasi plane-wave holding beam drives the optical cavity containing a nonlinear medium. The injection of narrow laser pulses creates persistent localized intensity peaks in the output (cavity solitons). }
\label{fig:6}
\end{figure}

Figure \ref{fig:6} shows the procedure to generate cavity solitons using nonlinear optical resonators. A coherent and stationary holding beam is injected into the cavity, and in order to create solitons one injects also short and narrow writing pulses. Subsequently, the solitons can be erased individually by injecting again the writing pulses with an opposite phase. The possibility of writing and erasing cavity solitons, and also of controlling their position by means of phase and amplitude gradients in the holding beam, makes them interesting in view of possible applications \cite{Lugiato03,Ackemann09}.

Because of its simplicity, the LLE has been even called the \lq\lq hydrogen atom\rq\rq of nonlinear cavities \cite{Firth02a}. However, the first experimental observation of cavity solitons has been realized using broad area vertical-cavity surface-emitting semiconductor lasers below threshold \cite{Barland02}, which are theoretically described by a model more complex than the LLE, because it includes the time evolution of the carrier density of the semiconductor \cite{NLOS}.

\subsection{The temporal/longitudinal version of the LLE}\label{sec:2c}

\begin{figure}[ht]
\includegraphics[width=1.0\columnwidth]{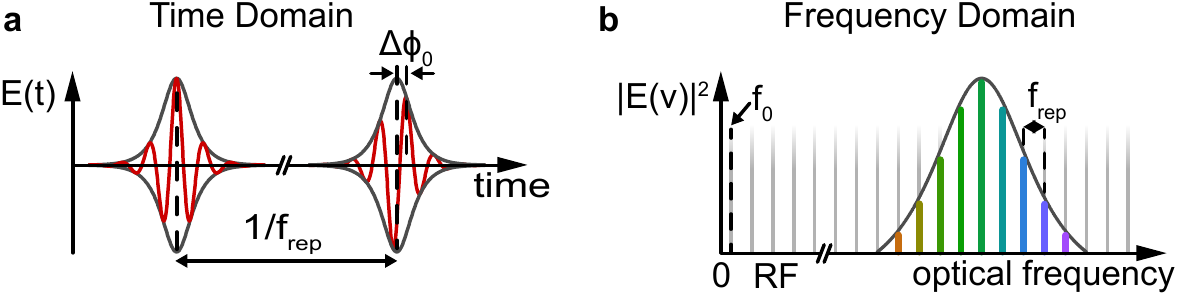}
\caption{\label{fig:CombsGeneral}Time and frequency domain picture of mode-locked
frequency combs. A periodic train of pulses with a
repetition rate $f_{\mathrm{rep}}$ (panel a) correspond to an optical frequency comb
spectrum of equidistant lines in the frequency domain (panel b). The
line spacing is given by $f_{\mathrm{rep}}$. The offset $f_{\mathrm{0}}$
of the frequency comb spectrum relates to the carrier-envelope phase
shift $\Delta\phi_{0}$ between two consecutive pulses via $f_{0}=f_{\mathrm{rep}}\cdot\Delta\phi_{0}/(2\pi)$.}
\end{figure}

In formulating the temporal/ longitudinal version of the LLE the authors of \cite{Haelterman92} extended to the dissipative case of cavity solitons the analogy between two kinds of Hamiltonian solitons: spatial solitons and temporal solitons. The principal step was to replace the 2D transverse Laplacian that appears in Eq. (\ref{eq:2}) and describes diffraction with the second derivative with respect to the retarded time in the cavity which describes dispersion. 
We write the equation in a form basically equivalent to that of Refs. \cite{Chembo13,Godey14}, which is commonly used in the literature on Kerr frequency combs:
\begin{equation}\label{eq:7}
\frac{\partial  E} {\partial t} = F -E - i \alpha E + i |E|^2 E - i \frac{\beta}{2} \frac{\partial^2 E}{\partial \theta^2}\,,
\end{equation}
where the cavity is assumed to be a ring of radius $R$, the angle $\theta$ is defined as $\theta=z/R$ and 
$\beta= -\left({\cal L}/TR^2\right)\left(\partial v_g/\partial\omega\right)_{\omega=\omega_0}$, 
with $v_g$ being the group velocity of light. An important point is that, once solved Eq. (\ref{eq:7}) with periodic boundary conditions in the interval $-\pi\le\theta\le\pi$, in the solution $E(\theta,t)$, $\theta$ must be replaced by $(\theta-(v_g/R)t)$, which means that the spatial pattern rotates along the cavity with velocity $v_g$.                                                                                                                                                   In the case of anomalous dispersion $\beta<0$ Eq. (\ref{eq:7}) has the same form that Eq. (\ref{eq:2}) has in the 1D case. Therefore all results obtained from the transverse LLE (\ref{eq:2}) hold, with the appropriate changes in notations, for the temporal/longitudinal LLE 
(\ref{eq:7}). This equation can be applied both to the case of fiber ring cavities as in \cite{Coen13,Coen13a}, and to whispering-gallery mode microresonators discussed in \cite{DelHaye07,
Chembo13,Godey14,Matsko11,Lamont13,Coillet14,Okawachi11,Grudinin12,Herr14} and integrated microrings \cite{Johnson2012,Saha13,Jung2013,Pfeifle14,Brasch16,Gaeta16,Weiner16}.  It is appropriate to mention that the majority of literature on Kerr combs uses frequency domain picture, and not time domain (see Fig.\ref{fig:CombsGeneral}). This is a natural frame to describe the threshold of parametric oscillations, as well as combs with few comb lines.

Sinusoidal-like longitudinal/temporal Turing patterns, corresponding to the transverse patterns described by Eq. (\ref{eq:6}), i.e. continuous wave driven modulation instability, has been first experimentally observed in \cite{Coen99,Coen01}. In these experiments a long fiber loop cavity was employed, which can demonstrate modulation instability, provided high finesse, with a continuous wave pump laser. From a viewpoint of the underlying four wave mixing process, this corresponds to parametric oscillations. Parametric oscillations, i.e. modulation instability were observed in optical microresonators for the first time in 2004 \cite{Kippenberg2004a,Savchenkov2004c} by work from Caltech and JPL in Pasadena. The advance that allowed this process, deriving from the weak third order nonlinearity, to be observed were the high optical Q-factors of the cavity. In contrast to conventional lasers, the threshold power for parametric oscillations scales with $V/Q^2$ ($V$ is the effective mode volume), allowing therefore for substantial power reductions, making it possible to observe the effect with micro-Watt pump powers.
Parametric oscillations can also be used to generate optical frequency combs\cite{DelHaye07}. In 2007, it was shown that such parametric oscillations enable to generate an optical frequency comb. This work demonstrated a new method to generate combs directly from a continous wave laser via parametric frequency conversion. In this work a toroid microresonator \cite{Armani03} was used, and the equidistance proven by using a fiber laser frequency comb to establish the equidistance between the widely (THz) spaced comb teeth.

\begin{figure}
\centering
\includegraphics[width=0.7\linewidth]{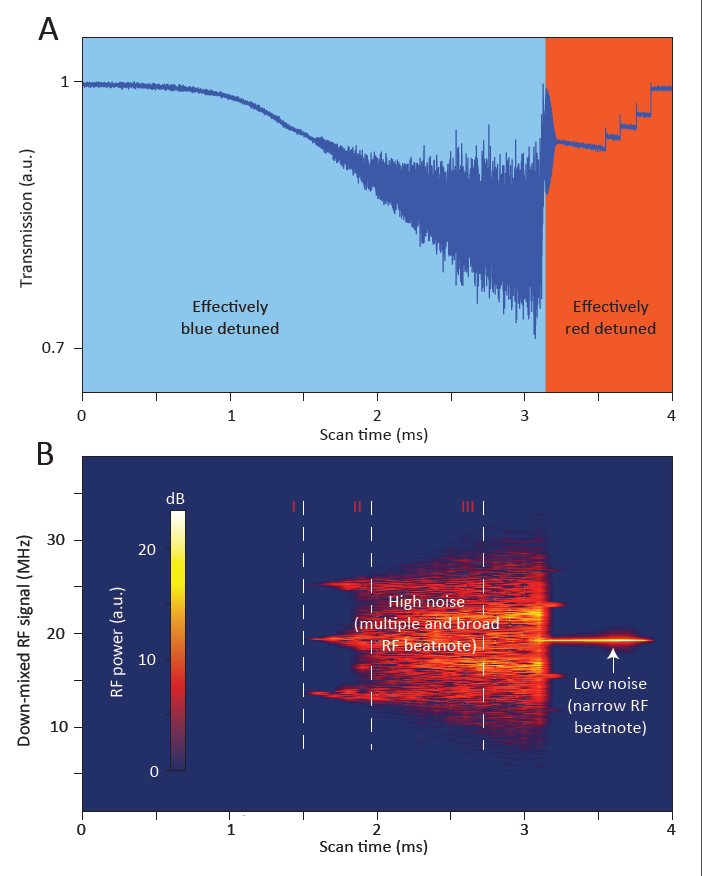}
\caption{Panel (A): First observation of DKS formation in the laser scan across a crystalline resonator. The effectively blue detuned side  (measured via a Pound-Drever Hall error signal) corresponds to modulation instability (MI), and gives rise to low coherence combs in the case of crystalline microresonators, as witnessed by a repetition rate that exhibits broad beat notes and satellite peaks, as shown in Panel (C). Strikingly, when detuning to the effectively red detuned side, which is conventionally unstable due to the thermal effect, a series of discontinuous steps occur, which have equal heights. These steps as has later been shown via simulations and experiments correspond to different number of DKS in the microresonator. Figure adapted from T. Herr, Nature Photonics, 2014.}
\label{fig:fig_HerrStepRF}
\end{figure}

In later, independent work, using macroscopic bulk optical fiber cavities, it was possible to enter the regime where cavity solitons are supported. This was established in Ref. \cite{Leo10} by using two pump fields: a strong CW pump laser as well as short optical pulses, that seeded the soliton formation In this manner a temporal cavity soliton (now typically referred to as dissipative Kerr soliton, DKS) was generated\cite{Leo10}. 
Using fiber cavities, also allowed observing breathing temporal solitons \cite{Leo11}, 
and temporal tweezing of solitons \cite{Jang15b}. Pattern formation in fiber-ring resonators is analyzed also in \cite{Schmidberger14}. 
This early work on fiber cavities also speculated that Kerr combs in microresonators may actually be DKS - which however is not the case. Indeed, early generated comb spectra corresponded to early stationary modulational instability (MI) states also known as periodic Turing patterns \cite{Chembo13,Godey14} that appear at effective blue detuning of the laser from resonance. In contrast, as is now well understood, DKS only form for the laser being effectively red-detuned. Such red detuned operation is conventionally unstable in optical microresonators due to the thermal effect \cite{Carmon04}. DKS were discovered and stably generated only in 2012 in micro-resonators by the work of T. Herr\cite{Herr14}, by overcoming the associated thermal instabilities. In this work it was observed that when scanning across the resonance of a crystalline optical microresonators, a series of discrete steps appeared (see Fig. \ref{fig:fig_HerrStepRF}). 

Experiments proved that these discontinuous, and remarkably regular discrete transmission steps, occurred on the red-detuned side of the cavity resonance, and were associated with a low noise comb state. Early work demonstrated the coherence via transient measurements of the comb's beat note. Using optimized laser scans it was possible to generate continuously circulating dissipative Kerr solitons, corresponding to femtosecond pulses with repetition rates in the GHz to THz regime. These developments are reviewed in the next section.



\section{Kerr frequency combs}

\begin{figure}[ht]
\center{\includegraphics[width=1.0\linewidth]{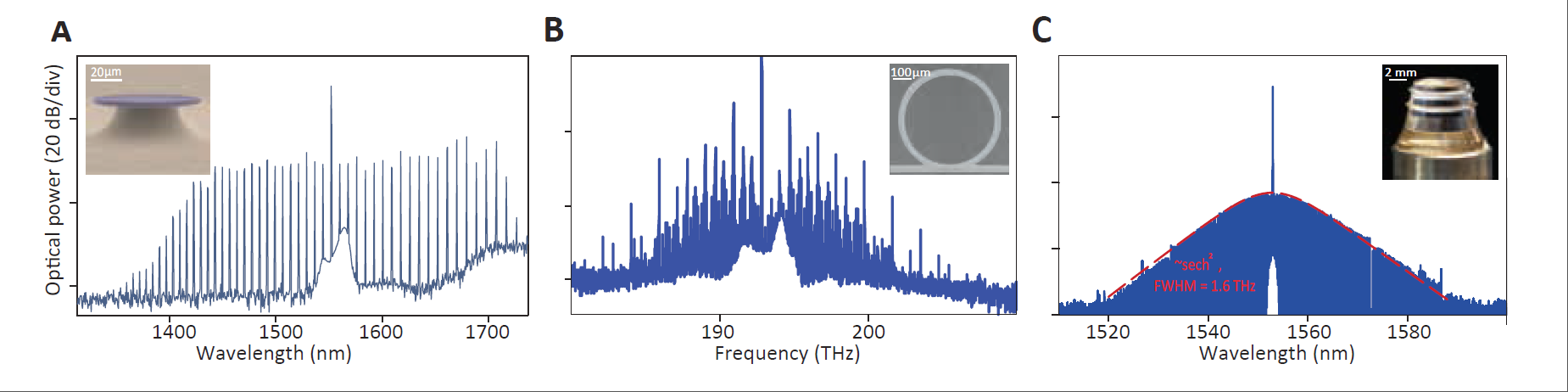}}
\caption{\label{fig:Exp_sols} Evolution of Kerr frequency comb generation in microresonators, as studied by two of the authors (TJK, MG). Panel (A) shows the first demonstration of an optical frequency comb in an toroid microresonator, as demonstrated in Del Haye, Nature, 2007. This work demonstrated an equidistant optical frequency comb generated from an ultra high Q toroid microresonator  via parametric interactions. The work demonstrated the equidistance to 1 part in $10^{17}$ by comparison with a fiber laser frequency comb. Panel (B) shows the spectrum in which a transition from high to low phase noise was observed in a (low quality factor) $Si_3N_4$ photonic integrated optical microresonator (T. Herr, Nature Photonics, 2012). This work showed that the high noise states could be overcome. Panel (C) shows the first spectrum of a stably accessed dissipative Kerr soliton in a crystalline resonator, which formed from a low coherence Kerr frequency comb.}
\end{figure}

The statement of Arthur Clarke that "any sufficiently advanced technology is indistinguishable from magic" can be related to microcomb technology as well. The formation of femtosecond optical solitons, corresponding to ultrastable broad frequency combs, self-organized from chaotic parametric oscillations in a continuous-wave driven micro-scale resonator, containing various imperfections of its mode structure, arguably can be perceived as a miraculous effect.
In particular, when taking into account the almost indistinguishable real experimental spectra from idealized theory and simulations 
\cite{Grudinin17}.
(see Fig.\ref{fig:Exp_sols} as an example) 
This perfect matching and ease of experimental verification provide an intriguing playground for the theory, forcing many scholars specializing in nonlinear optics and solitons to join this area. For experimentalists and applied physics, accustomed that high precision especially for metrological applications usually means bulky and heavy equipment, solitons in microresonators provide a promising alternative.  

\begin{figure}
\centering
\includegraphics[width=1\linewidth]{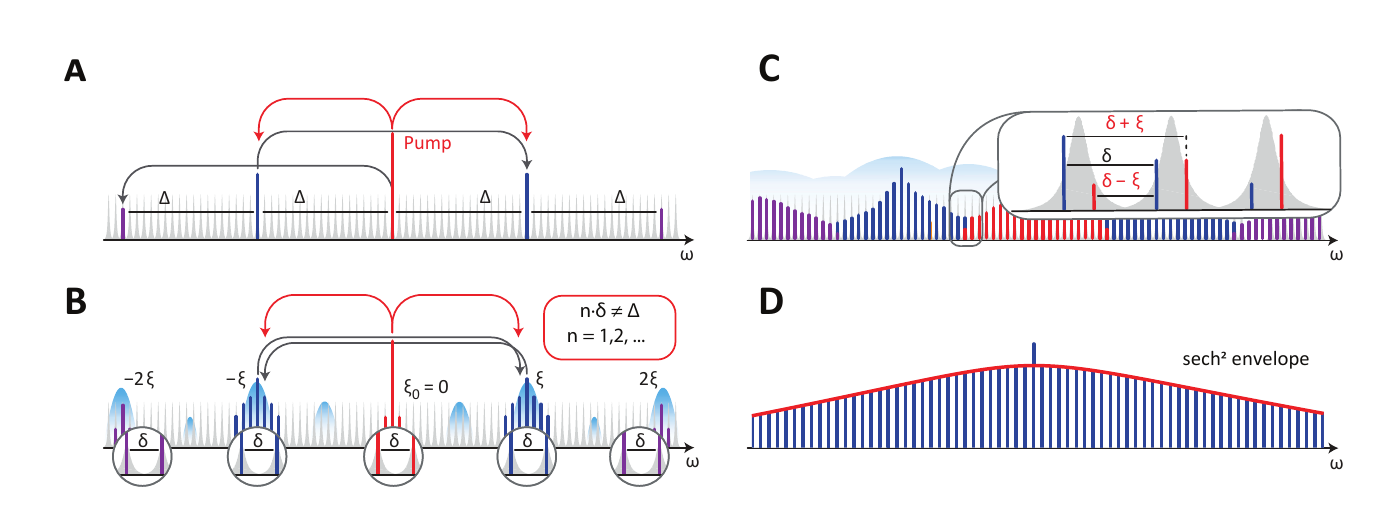}
\caption{Subcomb formation in optical microresonators.  Panel (A) shows formation of primary sidebands, that cover several free spectral ranges from the pump laser. Panel (B) show the formation of secondary sub-combs via a combination of non-degenerate four wave mixing processes. This produces sub-combs with the same repetition rate, but different off-set frequencies. Panel (C) shows how subcomb formation can lead to the situation where more than one comb line occupies a cavity resonance, leading to a high noise repetition rate beat note. Panel (D) shows how the low coherence state in the previous panel can undergo a transition and form a single DKS soliton state. }
\label{fig:fig_subcomb_tjk}
\end{figure}

High quality-factor (Q) optical microresonators with whispering-gallery modes since their discovery by Braginsky and co-workers in 1989 \cite{Braginsky1989} have long been considered promising for their nonlinear optical properties. The third order Kerr-nonlinearity, in particular, can initiate parametric oscillations in an optical microresonator. In this four-wave mixing process two pump photons are converted to one signal and one idler photon. This parametric conversion (sometimes also called hyper-parametric to distinguish it
from better known three-photon parametric oscillations associated with the second order optical nonlinearity), though being described in nonlinear optics long ago \cite{Klyshko1988},
has been observed in ultra high-Q toroidal fused silica and crystalline
resonators only in 2004\cite{Kippenberg2004a,Savchenkov2004c}. In these experiments the pump generated signal and idler sidebands. This process was accessed at very low threshold powers, which is explained on grounds that the threshold power for parametric oscillations scales as  $V_\mathrm{eff}/Q^{2}$, where $V_\mathrm{eff}$ is the effective mode volume. This is in contrast to conventional lasers.
In 2007 it was discovered that parametric oscillations in optical microresonators \cite{DelHaye07} give rise to optical frequency combs (see Fig. \ref{fig:Exp_sols}). By bridging the large (THz) mode spacing of the microresonators comb with a conventional fiber laser frequency combs, initial work demonstrated that "Kerr" frequency comb generated in optical microresonators are equidistant to 1 part in $10^{17}$, proving thus unambiguously the comb structure. 
These experiments therefore demonstrated that a frequency comb can be generated with the Kerr nonlinearity and FWM, and broke with the dogma that optical frequency combs require a pulse forming mode locking mechanism.  Moreover these experiments revealed that, contrary to the widely held expectation (based on modulator frequency combs), the phase noise does \emph{not} cascade and does not degrade coherence in the spectral wings of the combs.  The surprising cascading and proliferation of excited modes associated with FWM was demonstrated to lead to a broadband series of optical lines equidistantly spaced in the optical frequency domain, i.e. an optical "Kerr" frequency comb.  Shortly thereafter, it was moreover shown that such Kerr combs can also operate with electronically detectable repetition rates in the microwave domain \cite{Del'Haye2008}, and both carrier envelope offset frequency and mode spacing can be controlled via the pump laser and pump power, thereby offering full control over the comb spectrum.
In the subsequent years, Kerr-combs were demonstrated in progressively growing number of micro-resonator platforms, including crystalline resonators\cite{Savchenkov2008}, CMOS compatible integrated platforms with microring resonators such as silicon nitride ($\mathrm{Si_{3}N_{4}}$)\cite{Foster2011a,Levy2010,Razzari2010},
Hydex glass\cite{Moss2013}, as well as aluminum nitride \cite{Jung2013}
or diamond\cite{Hausmann2014} or high-Q silica microdisks \cite{Vahala15}.  
However, it was observed that in experiments involving broadband comb formation, that  the coherence of the comb spectrum was lost \cite{Del'Haye2008}. This effect was studied in detail in SiN and crystalline resonators \cite{Herr12}, where it was shown that the loss in coherence is related to the comb formation process, specifically due to the formation of \emph{sub-combs}, as outlined in Figure \ref{fig:fig_subcomb_tjk}.

\begin{figure}
\centering
\includegraphics[width=1\linewidth]{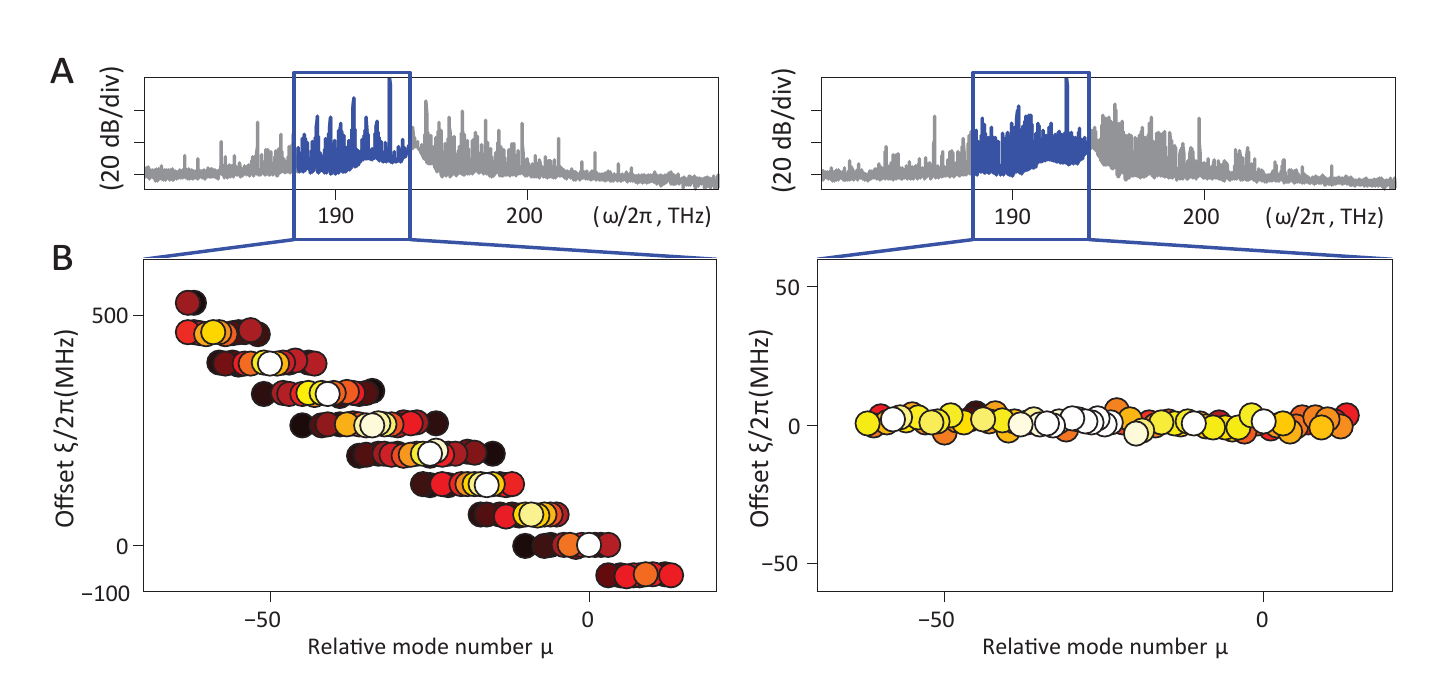}
\caption{Transition to low phase noise in an integrated optical SiN microresonator frequency comb (adapted from T. Herr, Nature Photonics 2012). Panel A shows the optical spectra of the generated Kerr frequency comb. Panel (B) shows the relative absolute optical frequency. In this representation a horizontal line corresponds to a equidistant optical spectrum. A transition of the comb is occurring that enforces the subcombs, that are initially exhibiting a different relative offset frequency (left panel) to synchronize and form a regular section of equally spaced comb lines (right panel).}
\label{fig:fig_lownoise_tjk}
\end{figure}

In case the mode-proliferation does not occur via FWM on sidebands close to the pump, the FWM leads to the formation of sub-combs. This subcomb formation was studied and observed experimentally, by reconstructing every absolute frequency of a SiN and crystalline Kerr frequency comb \cite{Herr12}. Although crystalline resonators feature exceptionally high Q, their low dispersion makes them susceptible to subcomb formation. Likewise the low Q factor of SiN microresonators leads to comb initiation via subcomb formation.
Pumping the resonator stronger will lead to the merging of subcombs, and the highly counter intuitive situation where more than one single comb line occupy a resonator mode. In this case the beat note exhibits multiplets, which can merge into a broad beat note. This behavior has been the case for virtually all optical microresonator combs, except comb formation in devices where the accumulated dispersion is sufficiently large to exceed the cavity decay rate, as in initial work by Del'Haye \cite{DelHaye07}. Indeed, many of the coherence criteria such as the repetition rate beat note, are not sufficient as indicator of a coherent comb state, as the pump laser dominates the beat note. Although not explicitly mentioned, many reported frequency combs are operating in the low coherence regime.
While these observations led the initial enthusiasm to cool down, it was found that there are states that despite subcomb formation can lead to coherent combs. One such states was reported in SiN and referred to as $\delta-\Delta$ matching, indicating that once the subcombs are separated by an integer multiple of the sub-comb spacing a low noise state can emerge.  The transition is show in Figure \ref{fig:fig_lownoise_tjk}.

This work showed that therefore coherence is attainable in platforms that are fully chip-integrated and proceed with sub-comb formation. Such low phase noise states were later also observed in silica disks \cite{Li2012} and in the case of SiN microresonators \cite{Saha13}, evidence of pulse formation observed.  

\begin{figure}[ht]
\center{\includegraphics[width=0.85\columnwidth]{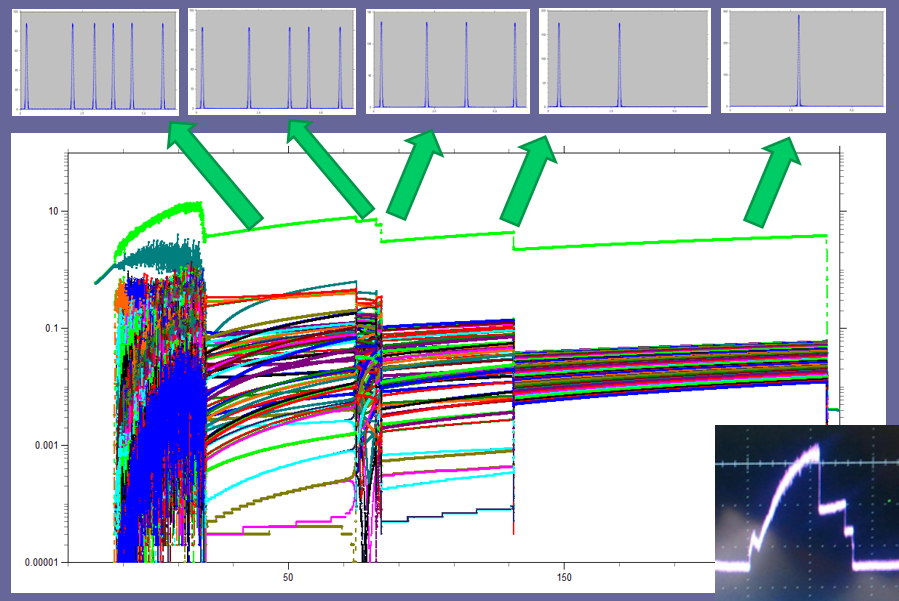}}
\caption{Surprising behavior of the microcavity resonance upon slow laser frequency scan. The internal power (light green) upper curve after reaching chaotic regime at large red detuning demonstrates step-like behavior. Each step corresponds to low noise soliton state with an integer number of solitons circulating round in the microresonator. Multiple color lines demonstrate magnitudes of different comb lines. The difference between the chaotic state and soliton states is clearly visible. Upper insets demonstrate corresponding waveforms calculated at time slices marked by green arrows. The figure is taken from the presentation \cite{BadHonnef_11} (2 Oct. 2011). The bottom-right photo inset -- is the first experimental observation of soliton steps (EPFL, 12 Aug. 2012)}
\label{fig:BadHonnef}
\end{figure}

It was early understood that in the mean field approximation the system is described by the LLE \cite{Matsko09,Chembo13}, which implies that 
that temporal dissipative solitons have to be present in microresonators \cite{Kaup78,Nozaki84,Wabnitz93,Barashenkov96}. Dissipative temporal solitons have been first generated and studied in analogous system with fiber loop cavity pumped by a continuous wave laser in 2010 by Leo et al. \cite{Leo10}. In these experiments however, in contrast to the microresonator case, the solitons were externally seeded and injected via a pulsed pump \cite{Leo10}. Although these experiments conjectured that Kerr comb formation may be due to dissipative soliton formation, such solitons were only observed several years later, and the original speculation turned out to be not correct as Kerr combs, in the form as those reported by Del'Haye, are Turing patterns, or stable MI combs.
An outstanding question in the study of Kerr combs was, however, how to achieve a dissipative soliton under continuous pumping. External seeding, as used in fiber cavities is challenging for microresonators with high repetition rate.
Though some publications claimed that chaotic Kerr frequency combs formed by CW pump may undergo DKS formation \cite{Herr12,Saha13}, methods to reach the single DKS state were only later developed. A surprising discovery were steps in the resonators transmission spectrum, which were shown in transient measurements to coincide with regions  of low phase noise, where the RF beat note collapses to a single narrow line. Keys answers of this behavior came from numerical simulations, that remarkably, predicted a similar step-like behavior. The first data, is shown in the figure \ref{fig:BadHonnef}.

Several approaches have been proposed for the theoretical analysis of parametric oscillations and  frequency combs \cite{Kippenberg2004a,Matsko2005a,Agha09,Chembo10PRA,Matsko09,Matsko11}. Besides description in the time domain the modal expansion approach in the frequency domain \cite{Chembo10PRA,Herr15,Hansson14} was found to be particularly useful in the context of microresonators as it allows easily to take into account particularities in the mode structure, such as avoided mode crossings.

The modified system of coupled mode equations in a dimensionless form developed for \cite{Herr12}(see the Supplementary Information) reads:
\begin{align}
\frac{\partial a_\mu}{\partial \tau}&=-[1+i\alpha_\mu] a_\mu +i\sum_{\mu',\mu''} a_{\mu'} a_{\mu''} a^*_{\mu'+\mu''-\mu}+ \delta_{0\mu} F.
\label{simeqs}
\end{align}
Here $a_\mu$ can be interpreted as the slowly varying  amplitude of the comb modes close to the mode frequency $\omega_\mu$ and $\tau=\kappa t/2$ denotes the normalized time ($\kappa$ is the energy damping rate), $\delta_{0\mu}$ is the Kronecker delta. The quantity $\alpha_{\mu}=2(\omega_\mu-\omega_p- \mu D_1)/\kappa$ is a formal measure of detuning defined by the cold resonance frequencies $\omega_\mu$ and an equidistant $D_1=2\pi f_{rep}$-spaced frequency grid. In the dimensionless form all frequencies, detunings and magnitudes are measured in units of cold cavity resonance linewidth so that $|a_\mu|^2=1$ corresponds to the nonlinear mode shift of half a cold resonance width,  which also corresponds to both single mode bistability and degenerate oscillations thresholds. Though (\ref{simeqs}) is a modified version of the one derived from the Maxwell's equations \cite{Chembo10PRA} it is also directly related to the temporal  LLE (\ref{eq:7}) if only the second order dispersion is present $\omega_\mu=\omega_0+D_1\mu+\frac{D_2}{2}\mu^2$ \cite{Chembo13} and may be considered as its discrete Fourier transform.

The numerical simulation using (\ref{simeqs}) of a comb formation in a dynamical regime, corresponding to usual experimental protocol, when $\alpha(t)$ is slowly varied from blue ($\alpha<0$) to red ($\alpha>0$) detuned regime, endeavored by one of the authors (MLG) revealed a surprising behavior (Fig.\ref{fig:BadHonnef}).
After the pump frequency is tuned further after the comb is formed, the internal power drops but not to the very low background level (as on Fig. \ref{fig:2}) but on an intermediate growing flat steps, characterized by an instant reduction of noise. These steps were identified as corresponding to an integer number of circulating solitons. These numerical observations were in agreement with experimental observations in a crystalline $MgF_2$ resonator.

\begin{figure}[ht]
\center\includegraphics[width=0.7\columnwidth]{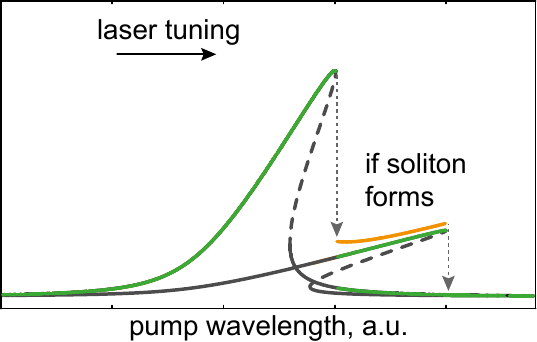}
\caption{The Kerr-nonlinear resonance can be described by bistability curves where the upper 
branch solution corresponds to high and the lower branch solution to low
intracavity power (see also Fig. \ref{fig:2}). When tuning into the resonance with decreasing optical frequency (increasing wavelength) the intracavity power follows
the upper branch of the Kerr-bistability curve. After the transition
to a soliton state the major fraction of the pump light is described
by the lower branch of the bistability curve. The fraction of the
pump light that propagates with the soliton inside the microresonator
experiences a larger phase shift and is effectively blue detuned on
the upper branch of another bistability curve. The extent of the \lq\lq soliton bistability curve" towards longer wavelength depends on the peak power of the solitons (i.e. the maximal nonlinear phase shift), the relative
height of the curve depends on the relative fraction of the pump light
that is affected by the high peak power soliton. The overall intracavity
power can be inferred by adding the bistability curves resulting in
the black curve.}
\label{fig:Stability-of-soliton-Curve}
\end{figure}

The emergence of the steps may be explained by referring to the cavity bistability curve that describes the power inside the cavity in dependence of the pump laser detuning. When approaching the cavity resonance from the blue detuned
side (laser frequency is higher than the resonance frequency and correspondingly
laser wavelength is shorter than the resonance wavelength) the intracavity
power will increase. This increase of intracavity power will then,
due to the Kerr-nonlinearity, effectively shift the resonance frequency
towards longer wavelength. As the laser detuning is further decreased
the intracavity power will steadily increase (upper branch on the
bistability curve), until the highest possible power is reached at
the point of effective zero-detuning. Beyond this point (now effectively
red-detuned) the intracavity-power will steeply drop to a much lower
value (lower branch on the bistability curve) as the Kerr-nonlinear
resonance shift vanishes (cf. Figure \ref{fig:Stability-of-soliton-Curve}).

Exactly at this transition to red detuning temporal DKS can emerge from the noisy intracavity waveform. It is important to note that the differentiation between red and blue detuning was not made in the mathematical or numerical simulation literature on dissipative Kerr solitons, nor on experiments with fiber cavities, but only made later with the experimental observation of DKS. In these experiments measurements of the effective detuning were made from the Kerr nonlinearity shifted resonance, via a Pound-Drever-Hall error signal, which enabled differentiating the bi-stable regime (cf. Figure \ref{fig:fig_HerrStepRF}).

In this case the formation of a high peak power soliton pulse will cause an additional Kerr-frequency shift such that the part of the pump light that sees the soliton pulse (i.e. co-propagating with the
soliton inside the cavity) follows now a second bistability curve (figure \ref{fig:Stability-of-soliton-Curve}). The result is a second power
drop at a larger detuning when the soliton vanishes. The transition from chaos to solitons is illustrated in Fig.\ref{fig:2Sols} simulated with LLE, which clearly demonstrates all the phases of this transition upon laser frequency tuning.
\begin{figure}[ht]
\center{\includegraphics[width=1\linewidth]{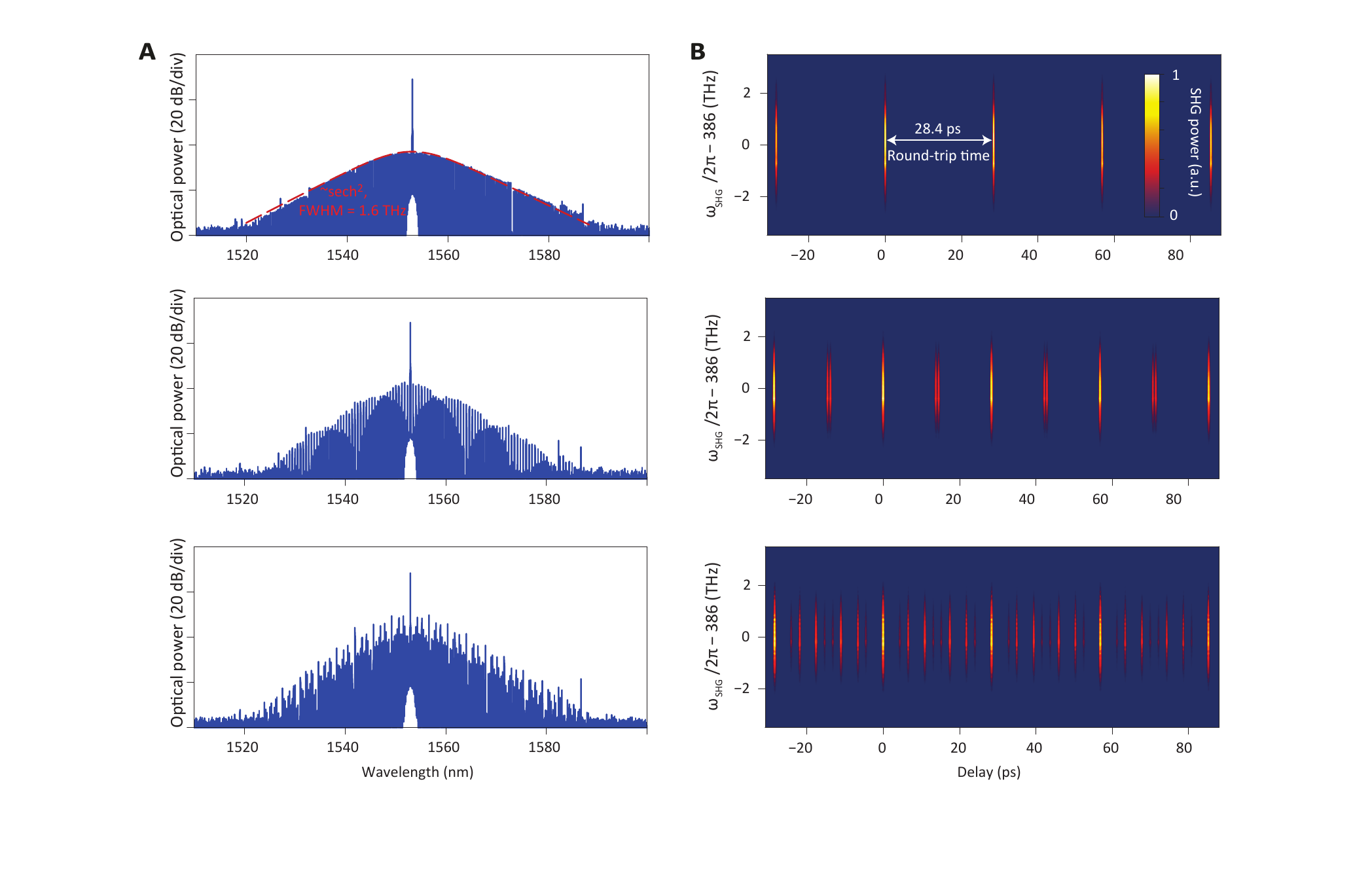}}
\caption{\label{fig:Exp_sols1} Experimentally observed spectra and corresponding temporal traces obtained with the FROG (frequency resolved optical gating) technique in MgF$_2$ microresonators for 1,2 and 3 dissipative Kerr soliton states. Figure adapted from \cite{Herr14}. Note the sech$^2$ soliton fit for the single soliton spectra, covering three orders of magnitude in relative optical power. For multi-soliton spectra the characteristic interference is observable giving rise to the patterns in the optical spectrum.}
\end{figure}

This surprising finding stimulated the two authors (MLG and TJK) with T. Herr to intensify the quest for DKS in crystalline MgF$_2$ microresonators, following the observations of the idealized numerical simulations. The first steps were observed a year later (photo inset in Fig.\ref{fig:BadHonnef}) and convincing spectral and temporal profiles were obtained (figure \ref{fig:Exp_sols1}). To demonstrate that the comb is indeed coherent, initial work focused on recording the transient beat notes, as show in figure 7. A key experimental technique that confirmed that operation in the regime where solitons can be supported, was the addition of a Pound-Drever-Hall error signal, that was added to the pump laser. This enabled to determine, during the laser scan, the effective laser detuning, i.e. the laser detuning from the thermally and Kerr shifted resonance. In these experiments the low noise regime was revealed to occur on the effectively red-detuned side of the cavity resonance. Access to the soliton state in steady state was compounded however by thermal effect\cite{Carmon04}, as red-detuned operation is not thermally stable. To nevertheless stably access the DKS state an optimized laser tuning scheme\cite{Herr14}, and later "power kicking"\cite{Brasch16} was developed that allowed overcoming the rapid temperature changes, associated with the discontinuous steps, and with the fact that the strong pump laser is tuned of the unstable, red-detuned side of the cavity resonance. Fortuitously, if this laser ramp is applied the soliton can be stably accessed: indeed, in the presence of a DKS the stability can be restored, as the soliton is a solution of the upper branch, being effectively blue detuned (and thus thermally stable\cite{Carmon04}), while the continuous wave pump laser is in the unstable red-detuned region, but far detuned from cavity resonance.  Applying this novel laser detuning technique allowed to stabilize soliton states and to prove unequivocally the observation of combs that are associated with soliton femtosecond pulses \cite{Soliton_Arxiv,Herr14}. This work also accessed all the important DKS states. 

\begin{figure}[ht]
\center{\includegraphics[width=1.0\linewidth]{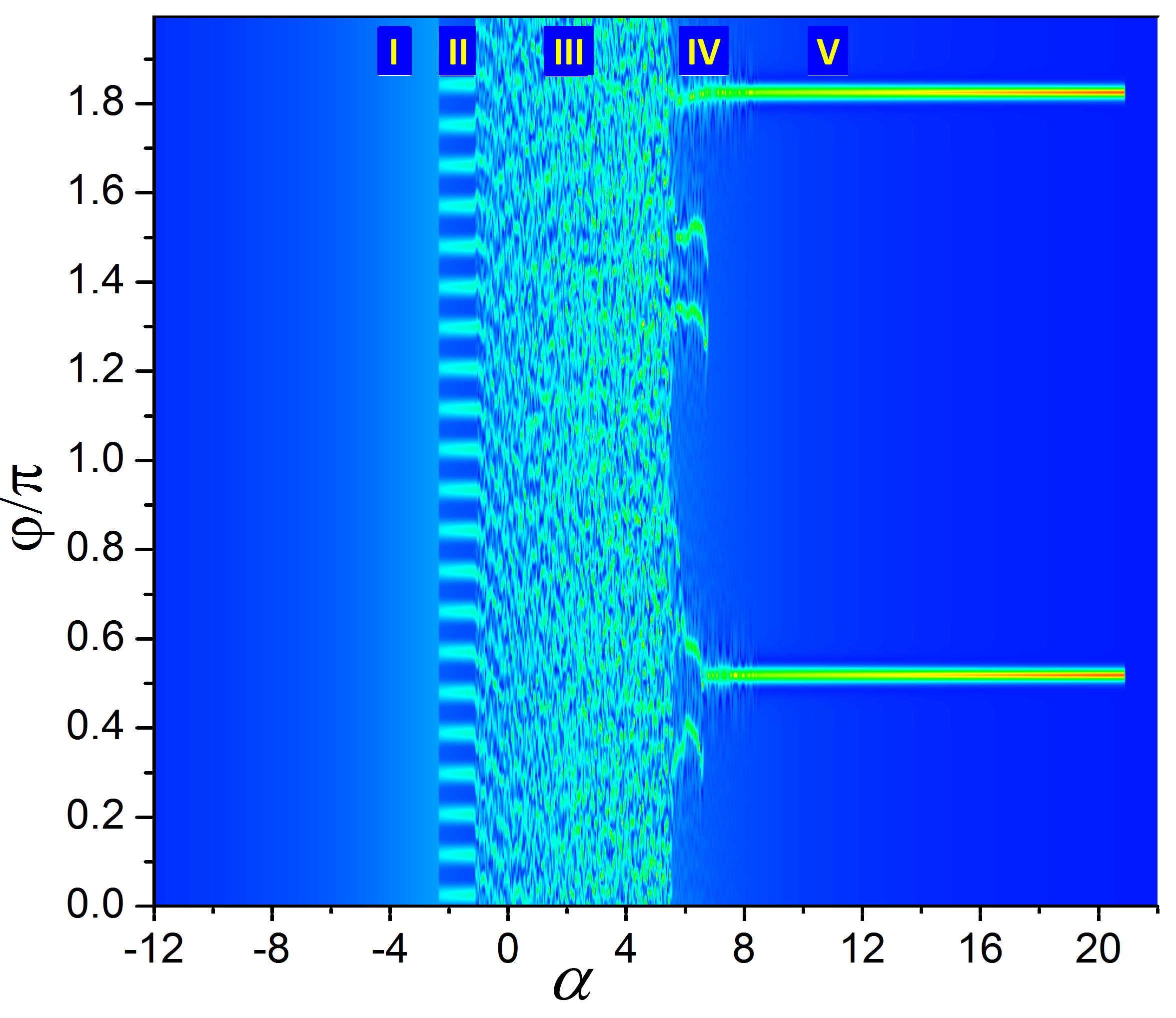}}
\caption{\label{fig:2Sols} From chaos to solitons. Internal angular intensity distribution in a Kerr microresonator when the detuning is gradually changed from blue to red: I -- no comb, only CW power at pump frequency, II -- Turing pattern, associated with primary comb formation, III -- chaotic regime associated with secondary combs formation and interaction, IV -- switching to soliton state with interacting breathers, V -- two stable dissipative Kerr solitons.}
\end{figure}

\section{Dissipative Kerr Solitons in Photonic Integrated microresonators}
Upon demonstration of DSK in crystalline resonators, an immediate outstanding and open question was, to which extent stable DKS could also be accessed in photonic integrated $\mathrm{Si_3N_4}$ microresonator, in which comb generation with octave spanning bandwidth (but low coherence) had been demonstrated.  Photonic integrated resonators are promising for actual applications, due to the ability to manufacture the devices wafer scale, the ability to chip integrated higher level of optical or electrical functionality and to allow integration with on chip lasers. Yet, SiN resonators exhibited lower quality factor and also significantly more irregularities in the resonators' dispersion. Such avoided mode crossing induced effects have been observed to hinder access of DSK formation in crystalline resonators \cite{Herr14PRL}. An open question was therefore if the significantly higher powers, along with resonator mode dispersion imperfection allow access to DSK states in photonic chip based microresonators. A key development to observing DKS in photonic chip based microresonators was the development of dispersion profiles with minimal influence of avoided mode crossing. However even in these cases, access to the soliton state was compounded by the very short duration (ns) of the soliton steps. 

\begin{figure}
\centering
\includegraphics[width=1\linewidth]{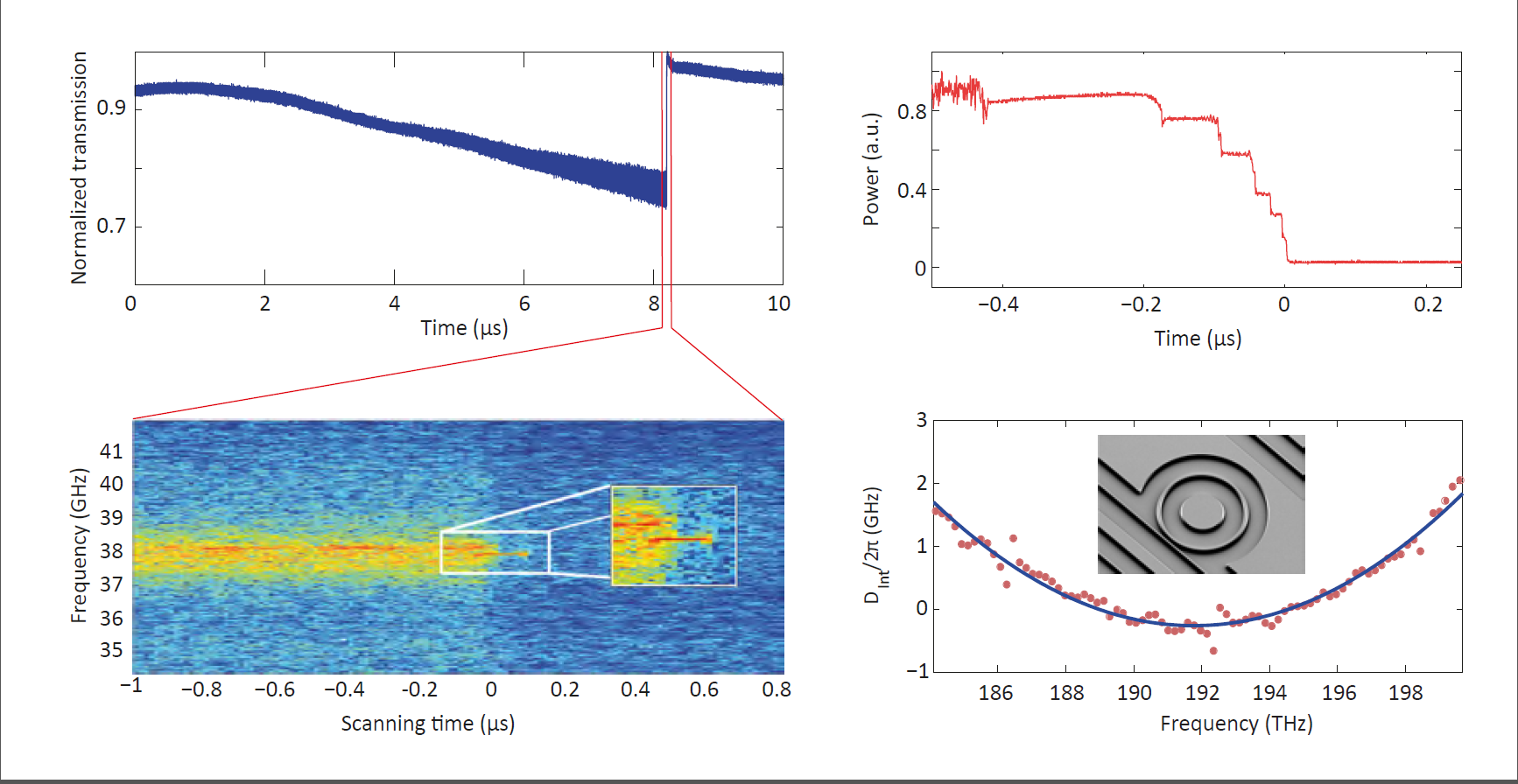}
\caption{First experimental evidence for DKS formation in photonic chip based microresonator (January, 2015). Panel (A) shows the transmission past a $\mathrm{Si_3N_4}$ microresonator, revealing very short steps (Panel B). The RF beat note associated with these steps show a characteristic collapse, indicative of a low noise comb state. Panel (D) shows the dispersion of the resonator, revealing anomalous GVD with little influence of mode crossings. Figures adapted from Brasch et a. Science 2016.}
\label{fig:figa}
\end{figure}

\begin{figure}
\centering
\includegraphics[width=1\linewidth]{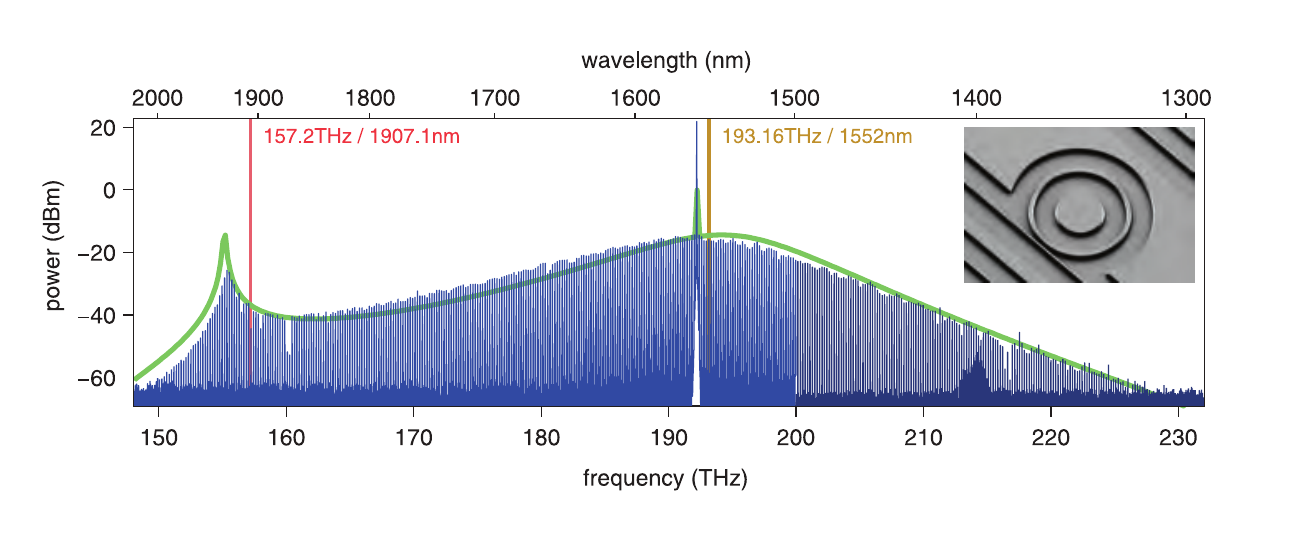}
\caption{Single DKS formation in a photonic chip based $\mathrm{Si_3N_4}$ optical microresonator. Figure adapted from Brasch et al. Science 2016.}
\label{fig:figc}
\end{figure}

The first DKS comb state was accessed, along with the observation of Soliton Cherenkov radiation, i.e. a dispersive wave, almost 2 years later (see Fig. \ref{fig:figa}). The access of the soliton required a new method "power kicking" to be developed, as well as samples with significantly improved resonator dispersion, i.e. the absence of avoided mode crossings \cite{Brasch16}. In this work it was possible to generate 2/3 of an octave from a 1550 nm pump laser, and in particular observe the formation of a dispersive wave. The theory and experiment showed excellent agreement, except for the absence of a dispersive wave recoil, that was compensated by the later understood Raman self frequency shift.
Upon stable access of the DKS state in SiN, the results were extended in other integrated platforms. Today soliton micro-combs are routinely obtained in many other laboratories and in myriad of resonator platforms. \cite{Vahala15,Liang15,Grudinin15,Gaeta16,Weiner16}.
Of all platforms SiN has witnessed remarkable advances in the last years, and enabled photonic chip-scale frequency combs with the largest level of on chip integration (see Fig. \ref{fig:figc}). Moreover, SiN combs have been applied to optical frequency synthesis, coherent communications, ultrafast distance measurements and spectrometer calibration. Most recently battery powered, on chip DKS combs have been demonstrated, highlighting the technological readiness level.

\section{Conclusion}
In this paper we have outlined the research history of two topics, namely the LLE and broadband Kerr frequency combs, that are related by an intrinsic link. From a historical viewpoint, the relation is not of the kind which links a cause to an effect, because the experimental discovery of Kerr combs was realized without the awareness and connection to the LLE. The observation of DKS however have linked these two research topics, and and has given major significance to the LLE within the context of soliton Kerr frequency combs, and likewise has given a new highly relevant technological application scenario to the field of pattern formation, which is uncommon in the vast area of pattern formation, in which theoretical and experimental investigations have typically been of purely fundamental character. From the perspective of soliton micro-combs, temporal pattern formation is the underlying Physics that has brought major advances, enabling soliton micro-combs to be used for dual comb spectroscopy \cite{Suh16}, counting the cycles of light \cite{Jost15},  terabit per second coherent communication at the sender and receiver side \cite{Marin17}, soliton based ultrafast distance measurements \cite{Suh18,Trocha18} as well as astrophysical spectrometer calibration for exo-planet searches \cite{Obrzud17,suh2018searching} and even make fully photonic integrated frequency synthesizers \cite{Spencer17}. Soliton microcombs have the potential to make frequency comb technology portable, chip integrated and sufficiently compact to be widely deployed.
This example shows that sometimes the scientific evolution is not linear but complex, and it is in a sense reassuring to see that some seeds are not lost but mature over long periods of time, when the technological progress allows, because  science is essentially  a collective process, and proceeds as a river flow which  transports  knowledge in ways that it is impossible to identify while it occurs, but can only be recognized a posteriori. 
It is also interesting to observe that the criterion of simplicity, followed in the construction of the LLE,  has turned out to be visionary with respect to achievements  realized two decades later. 

\vskip6pt

\enlargethispage{20pt}




\end{document}